\documentclass[aps,final,notitlepage,oneside,onecolumn,nobibnotes,nofootinbib,
superscriptaddress,noshowpacs,centertags]{revtex4-1}

\usepackage[english]{babel}
\usepackage{graphicx}
\usepackage{latexsym}
\usepackage{amssymb}
\usepackage{amsmath}
\usepackage{float}

\begin{document}

\title{Formation and clustering of primordial black holes in Brans--Dicke theory}
\author{V. A. Berezin}\thanks{e-mail: victor.a.berezin@gmail.com}
\affiliation{Institute for Nuclear Research of the Russian Academy of Sciences, Moscow, 117312 Russia}
\author{V. I. Dokuchaev}\thanks{e-mail: dokuchaev@inr.ac.ru}
\affiliation{Institute for Nuclear Research of the Russian Academy of Sciences, Moscow, 117312 Russia}
\author{Yu. N. Eroshenko}\thanks{e-mail: eroshenko@inr.ac.ru}
\affiliation{Institute for Nuclear Research of the Russian Academy of Sciences, Moscow, 117312 Russia}
\author{A. L. Smirnov}\thanks{e-mail: smirnov@ms2.inr.ac.ru}
\affiliation{Institute for Nuclear Research of the Russian Academy of Sciences, Moscow, 117312 Russia}

\date{\today}

\begin{abstract}
The formation of primordial black holes in the early universe in the Brans-Dicke scalar-tensor theory of gravity is investigated. Corrections to the threshold value of density perturbations are found. Above the threshold, the gravitational collapse occurs after the cosmological horizon crossing. The corrections depend in a certain way on the evolving scalar field. They affect the probability of primordial black holes formation, and can lead to their clustering at large scales if the scalar field is inhomogeneous. The formation of the clusters, in turn, increases the probability of black holes merge and the corresponding rate of gravitational wave bursts. The clusters can provide a significant contribution to the LIGO/Virgo gravitational wave events, if part of the observed events are associated with primordial black holes.
\end{abstract}

\maketitle 

\tableofcontents


\section{Introduction}

The principal possibility of the primordial black holes (PBHs) formation in the early universe was stated in the works~\cite{ZelNov67,Haw71}. Several mechanisms were proposed, among~them: collapses of adiabatic density perturbations~\cite{CarHaw74,Car75}, collapses of perturbations at early dust-like stages~\cite{KhlPol80, ZabNasPol87}, collapses of domain walls~\cite{BerKuzTka83,Khl98,KhlRubSak04}, and collapses of baryon charge fluctuations~\cite{DolSil93,Dol18,DolPos20}. It is possible that some of the LIGO/Virgo events are explained by the merge of PBHs~\cite{Naketal97,Ioketal98,Sasetal16,Blietal16,Doletal16}. The~value of density perturbations that are necessary for the formation of a PBH can be achieved due to the presence of features in the inflationary potential~\cite{Sta92,IvaNasNov94}, as~well as in inflationary models with several scalar fields~\cite{Yok95}.

The process of PBHs formation depends on the underlying theory of gravity. In~addition to numerous analytical and numerical calculations that are performed in the framework of General relativity, there are studies of the PBHs formation in modified gravity theories. The~paper~\cite{BarCar96} considered the effect of the gravitational constant changes in the Brans-Dicke theory on the number of PBHs formed. The~evolution of PBHs population in the Brans-Dicke cosmology was also studied in~\cite{NayMajSin10,AliZar20}. A~PBH is only formed from density perturbation that is greater than the certain threshold value. The~number of PBHs is very sensitive to the threshold of their formation, and~in General relativity, the~threshold is $\sim$~$1/3-0.5$. For~some principal issues that are related to the threshold, see the paper~\cite{EscGerShe20}.  The~modification of the threshold in the Eddington-inspired-Born-Infeld gravity was investigated in~\cite{Che19}, and~it was shown that the threshold in this theory of gravity is not constant, but~it depends on the epoch of PBH~formation.

In this paper, we also calculate corrections to the threshold of PBHs formation, but~we do it within the framework of the Brans-Dicke theory. In~contrast to~\cite{BarCar96,NayMajSin10,AliZar20}, we write down equations for the evolution of spherical-symmetric perturbations in the Brans-Dicke theory and consider the gravitational collapse. The~corrections depend on the scalar field of the Brans-Dicke theory. This~dependence leads to the fact that PBHs in the regions with different scalar field are formed with different probability. The~scalar field modulates the large-scale distribution of PBHs. The~modulation effect is similar to the well-known biasing effect for galaxies. The~galaxies are more likely to form in regions with higher density than in regions with lower density~\cite{Kai84}.

In~\cite{Mes74,Mes75}, attempts were made to explain the long-wave spectrum of perturbations (on the scales of galaxies and clusters of galaxies) by statistical fluctuations in the PBHs number density. This mechanism is ineffective, as shown in~\cite{Car77}. In~\cite{Mes74}, the~initial (deterministic) long-wave fluctuations in a mixture of PBHs and radiation were considered. It was assumed that the perturbations in the PBHs have the same amplitude as the initial perturbations in the radiation. A~more precise approach requires accounting for the biasing effect that accompanies the formation of PBHs at the background of long-wave perturbations. The~possibility of the biasing for PBHs within the General relativity was proposed in~\cite{Chi06}. However, later, by~more precise calculations, it was shown that this effect is very small~\cite{DesRio18, SuyYok19}. The~effect, in~a sense that is similar to the biasing effect, can also appear due to the non-Gaussian nature of initial perturbations~\cite{ByrCopGree12,TadYok15,YouByr15}.

A working mechanism for PBHs clustering was developed in~\cite{KhlRubSak04}, where it was shown that, for a certain inflationary potential, spherical domain walls are formed, which break up into PBHs. The~resulting PBH clusters may have a number of observational consequences for gravitational-wave astronomy and the formation of early galaxies and quasars~\cite{Bel19}. The~influence of PBH clustering on the rate of their merge within the General relativity was discussed in~\cite{AtaSanTri20}.

In this paper, we investigate the new biasing effect for PBH in the Brans-Dicke theory. We show that the biasing (clustering) of the PBHs actually occurs. It can affect the rate of PBH merge, which may be one of the sources of gravitational waves detected by LIGO/Virgo.

The Brans-Dicke theory~\cite{BraDic61} is one of the most well-known extensions of General relativity, reducing to the General relativity in the limiting case when one of the parameters of the theory $\omega\to\infty$. Therefore, as~long as the experimental data do not exclude General relativity, the~Brans-Dicke theory will also remain viable with the appropriate choice of parameters. Currently, the~observational data provide the following restrictions on the Brans-Dicke parameter $\omega>40,000$ from the Cassini-Huygens~measurements.

Perturbations of curvature and perturbations of scalar field that lead to the formation and clustering of PBHs are generated at the inflation stage. In~this paper, we do not fix any particular theory of the inflation in scalar-tensor theory. We note only that, in~Brans-Dicke inflation theory, curvature perturbations and perturbations in a scalar field other than inflation can be independent, according to the theoretical calculations of~\cite{StaYok95,TahRezKar16}. This is enough to demonstrate the new mechanism for PBHs clustering. Although,~the specific type of inflationary perturbations and quantitative characteristics of the clustering depend, of~course, on~the specific inflation~theory.


\section{Primordial Black Holes in the Brans-Dicke~Theory}

\subsection{Basic~Equations}

We consider the Brans-Dicke theory with the Lagrangian
\begin{equation}
\mathcal{L}=\phi R-\frac{\omega g^{\mu\nu}\partial_\mu\phi\partial_\nu\phi}{\phi}+16\pi\mathcal{L}_m,
\label{lagr}
\end{equation}
where $\omega=const$ and~$\mathcal{L}_m$ is the Lagrangian of matter. In~calculations, we assume that the speed of light $c=1$.

\begin{figure}[H]
\centering
\includegraphics[width=6 cm]{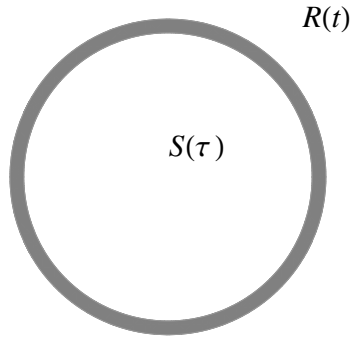}
\caption{The collapsing region is modelled by a part of a closed universe with the scale-factor $S(\tau)$. Outside the perturbed region the flat universe is assumed with the scale-factor $R(t)$. Additionally,~some transition layers are~possible.}
\label{gr1}
\end{figure}

Following the approach of~\cite{Car75}, we model a perturbed region of space collapsing to the PBH as a part of a closed universe with a metric
\begin{equation}
ds^2=d\tau^2-S^2\left(\frac{dr^2}{1-r^2}+r^2d\Omega^2\right),
\end{equation}
where $S=S(\tau)$ is the scale factor. Additionally,~the metric outside the perturbed region is the metric of the flat~universe
\begin{equation}
ds^2=dt^2-R^2\left(dr^2+r^2d\Omega^2\right),
\end{equation}
where $R=R(t)$, see Figure~\ref{gr1}. Between~the regions some transition intermediate layers are possible, see the discussion in~\cite{Che19}.
In contrast to~\cite{Car75}, we assume that the evolution is governed not by the General relativity, but~the Brans-Dicke theory.
In the Brans-Dicke theory, the~cosmological equations for a flat and closed cosmological models have the form~\cite{Wei72},
\begin{equation}
\frac{1}{R^2}\left(\frac{dR}{dt}\right)^2=\frac{8\pi}{3}\frac{1}{\bar\phi}\bar\varepsilon-\frac{\dot{\bar\phi}}{\bar\phi}\frac{\dot R}{R}
+\frac{\omega}{6}
\frac{\dot{\bar{\phi^2}}}
{\bar{\phi^2}},
\label{meqr}
\end{equation}
\begin{equation}
\frac{1}{S^2}\left(\frac{dS}{d\tau}\right)^2+\frac{1}{S^2}=\frac{8\pi}{3}\frac{1}{\phi}\varepsilon-\frac{\dot{\phi}}{\phi}\frac{\dot S}{S}
+\frac{\omega}{6}
\frac{\dot{\phi^2}}{\phi^2},
\label{meqs}
\end{equation}
where, in the Equation~(\ref{meqr}), a point means a derivative over $t$, and~in Equation (\ref{meqs}) a point means a derivative over $\tau$ (and further similarly).

We set the initial data at some early time $t_i$ ($\tau_i$) after the inflation stage, but~when all density perturbations are still small and described by linear theory. Values at the initial moment are marked with the index ``i''. The~cosmological epoch of radiation dominance is considered (equation of state $p=\varepsilon/3$); therefore, in the Equations~(\ref{meqr}) and (\ref{meqs}), one has
\begin{equation}
\bar\varepsilon=\bar\varepsilon_i\frac{R_i^4}{R^4}, \qquad \varepsilon=\varepsilon_i\frac{S_i^4}{S^4}.
\end{equation}

The scalar field evolves as $\dot{\bar\phi}R^3=const$, $\dot{\phi}S^3=const$ \cite{Wei72}, so
\begin{equation}
\dot{\bar\phi}=\dot{\bar\phi}_i\frac{R_i^3}{R^3}, \qquad \dot{\phi}=\dot{\phi}_i\frac{S_i^3}{S^3}.
\label{dotphi}
\end{equation}

Note that, for $\omega\dot{\phi}_it_i/\phi_i\ll 1$, the~last term on the right hand side Equation (\ref{meqs}) can be ignored throughout cosmological evolution, but, in general, the last term may be~important.

Additionally, the last closing equation
\begin{equation}
\frac{d\tau}{dt}=\frac{S}{R(1+\delta_i)^{1/4}}
\label{taut}
\end{equation}
gives a relation between $\tau$ and $t$ \cite{Har70}. This equation follows from the general expressions for the comoving coordinate system~\cite{LL-II} and it is valid both in General relativity and in Brans-Dicke theory. The~value of the density perturbation is defined as
\begin{equation}
\delta=\frac{\varepsilon-\bar\varepsilon}{\bar\varepsilon}.
\label{defdelta}
\end{equation}

In Equation (\ref{taut}), the~perturbation is taken at the initial moment $t_i$ ($\tau_i$), and~the initial constant-time hypersurface is chosen, so that, when $t=t_i$ ($\tau=\tau_i$), the~following conditions are met~\cite{Har70}
\begin{equation}
S_i=R_i, \qquad \left.dS/d\tau\right|_{\tau_i}=\left.dR/dt\right|_{t_i}.
\label{hyper}
\end{equation}

Note that, in other works~\cite{KopHofWel11,CarHar15}, where the formation of PBHs was considered in the framework of General relativity, the~constant-time hypersurface is chosen differently (we use the choice that coincides with the choice in~\cite{Car75,Har70}), and~it must be taken into account when comparing the results of the zero approximation of General~relativity.

In the following sections, we first find the solution of Equations (\ref{meqr}) and (\ref{meqs}) in the zero approximation at $\dot{\phi}=0$ (coinciding with the General relativity), and~we then solve the problem in the first linear approximation in Brans-Dicke~theory.

\subsection{Zero~Approximation}

In the zero approximation, $\dot\phi=\dot{\bar\phi}=0$ and $\phi=\bar\phi=1/G$, where $G$ is the Newtonian gravitational constant. This approximation matches the results that were obtained in General relativity in~\cite{Car75}. Here, we will present the known results from~\cite{Car75} and write down some new relations for the evolution of $\delta$ and for the threshold of PBH~formation.

In the zero approximation, we denote $A\equiv S^{(0)}$. The~condition $dA/d\tau=0$ presented in Equation~(\ref{meqs}) gives the scale factor at the moment $\tau_{\rm max}$ of the maximum expansion of the disturbed~region
\begin{equation}
A_{\rm max}=\frac{(1+\delta_i)^{1/2}A_i}{\delta_i^{1/2}}\approx\frac{A_i}{\delta_i^{1/2}},
\end{equation}
because $\delta_i\ll1$, and~from Equation (\ref{hyper}), we have
\begin{equation}
\delta_i=4G\phi_i\frac{t_i^2}{A_i^2}=4\frac{t_i^2}{A_i^2}.
\end{equation}

The evolution of the scale factor in the zero approximation is described by the equation
\begin{equation}
\frac{dA}{d\tau}=\left(\frac{A_{\rm max}^2}{A^2}-1\right)^{1/2}.
\end{equation}

Introducing the parameter $\psi$, one obtains the solution in the parametric form~\cite{Har70}
\begin{equation}
A=A_{\rm max}\sin{\psi},
\end{equation}
where the relation of the parameter $\psi$ to $\tau$ has the form
\begin{equation}
\tau=A_{\rm max}(1-\cos{\psi}),
\end{equation}
and the relationship between $\psi$ and $t$ is as follows
\begin{equation}
\psi=\arcsin\left(\frac{\delta_i^{1/2}}{(1+\delta_i)^{1/2}}\right)+\frac{\delta_i^{1/2}}{(1+\delta_i)^{1/4}}\left[\left(\frac{t}{t_i}\right)^{1/2}-1\right].
\label{psit}
\end{equation}

In the last expression, in~contrast to~\cite{Har70}, we took the next order corrections for $\delta_i$ into account, which will be required later in order to calculate $\delta(t)$.

The maximum expansion of the disturbed region corresponds to the moment of time
\begin{equation}
t_{\rm max}=t_i\left(\frac{\pi}{2}\right)^2\frac{(1+\delta_i)^{1/2}}{\delta_i},
\end{equation}
and at the same time
\begin{equation}
\tau_{\rm max}=t_i 2\frac{(1+\delta_i)^{1/2}}{\delta_i}.
\end{equation}

When selecting a constant-time hypersurface, the~ratio of densities at the time of maximum expansion is $\varepsilon/\bar\varepsilon=(\pi/2)^4$. In~other works (for example,~\cite{KopHofWel11,CarHar15}), a~different choice of the hypersurface was used, in~which the time in an undisturbed flat region of the universe is $t=\tau$, and~this choice results in $\varepsilon/\bar\varepsilon=4$, as~already mentioned above.

The value of the perturbation is expressed, as follows
\begin{equation}
\delta=\frac{\varepsilon-\bar\varepsilon}{\bar\varepsilon}=\left(\frac{t}{t_i}\right)^2\frac{\delta_i^2}{(1+\delta_i)}\frac{1}{\sin^4{\psi}}-1,
\label{deltapsi}
\end{equation}
where $\psi$ is related to $t$ by Equation (\ref{psit}).
For $t\ll t_{\rm max}$, one obtains
\begin{equation}
\delta\approx\frac{1}{3}\delta_i\left(\frac{t}{t_i}\right)^{-1/2}+\frac{2}{3}\delta_i\left(\frac{t}{t_i}\right)
\end{equation}

It means that, in this coordinate system, the~falling perturbation mode evolves as $\propto t^{-1/2}$, in~contrast to~\cite{CarHar15}, where the law $\propto t^{-1}$ was obtained. Additionally,~the evolution of a growing mode on scales larger than the horizon occurs according to the law $\propto t$ known for a synchronous reference~frame.

The PBH formation criterion was analytically obtained in~\cite{Car75}. The~region of space with $\delta>0$ was modelled as part of a closed universe (Friedman model). The~gravitational collapse of such a region with the formation of a PBH occurs if the relative value of the perturbation at the moment of horizon crossing $\delta_{\rm H}$ satisfies the following conditions:
\begin{equation}
\delta_c\le\delta_{\rm H}\le1.
\label{usl0},
\end{equation}

In early studies, criterion for the threshold $\delta_c$ was used, according to which PBH is formed, if~the radius of the disturbed region at the time of of its expansion stop exceeds Jeans radius. It corresponds to the left inequality in Equation (\ref{usl0}). Additionally,~the right inequality corresponds to the formation of a black hole, not a separate universe. As~noted in~\cite{HarYooKoh14}, this criterion contains ambiguity in the expression for the Jeans radius and can, therefore, only be considered as an estimation. Further, when considering the scalar-tensor theory, we will look for small corrections to the threshold of PBH formation, so, in this paper, we use the refined PBH formation criterion that was proposed in~\cite{HarYooKoh14}. According to the improved criterion, a~PBH is formed if the sound wave does not have time to reach the center from the periphery of the disturbed region by the time the expansion stops. In~this case, the~Jeans radius is assumed to be equal to the sound~horizon.

We write the metric of the perturbed region in the form
\begin{equation}
ds^2=d\tau^2-A^2\left(d\chi^2+\sin^2\chi d\Omega^2\right),
\end{equation}
then the equation of the sound wave has the form~\cite{HarYooKoh14}
\begin{equation}
A\frac{d\chi}{d\tau}=-\frac{1}{\sqrt{3}}.
\label{soundeq}
\end{equation}

Substituting the above expressions, we obtain
\begin{equation}
\frac{d\chi}{d\psi}=-\frac{1}{\sqrt{3}}.
\end{equation}

Integrating up to the moment of maximum expansion $\psi=\pi/2$, we have
\begin{equation}
\chi_{\rm max}=\frac{\pi}{2}\frac{1}{\sqrt{3}},
\end{equation}
which, in the case of the equation of state $p=\varepsilon/3$, matches the result obtained in~\cite{HarYooKoh14}.
However, we use a different choice of hypersurface, so the further calculation will be different.
The conditions for the Jeans radius (sound horizon) and condition for cosmological horizon crossing make up the system of equations
\begin{equation}
R_{\rm J}=A_{\rm max}\sin\chi_{\rm max}=A_{\rm max}\sin\chi_a,
\end{equation}
\begin{equation}
R_{\rm H}=2t_H=A_{\rm max}\sin\chi_a\sin\psi_{\rm H},
\end{equation}
where $\chi_a$ is the boundary of the disturbed region and~$\psi_ {\rm H}$ is the value of the parameter $\psi$ at the moment of the horizon crossing.
From these two equations, one obtains
\begin{equation}
2t_H=A_{\rm max}\sin\left(\frac{\pi}{2}\frac{1}{\sqrt{3}}\right)\sin\psi_{\rm H}.
\label{syreq}
\end{equation}

In the limit $\delta_i\ll1$, the Equation (\ref{syreq}) is transformed into the following nonlinear equation
\begin{equation}
\xi=\sin(\xi^{1/2})\sin\left(\frac{\pi}{2}\frac{1}{\sqrt{3}}\right),
\label{nonlineq}
\end{equation}
where
\begin{equation}
\xi=\frac{\delta_it_H}{t_i}.
\end{equation}

Numerical solution gives the root of the above equation
\begin{equation}
\xi=\xi_0\simeq0.52.
\end{equation}

Substituting $t_H=\xi_0t_i/\delta_i$ in Equation (\ref{deltapsi}), we obtain the threshold for the PBH formation in the zero~approximation
\begin{equation}
\delta_c^{(0)}=0.42.
\end{equation}

Note that, in other coordinate systems, the~threshold value is~different.

\subsection{Corrections to the Black Hole's Formation~Threshold}

In the following approximation, we take the evolution of the gravitational constant (change of the field $\phi$) in the Brans-Dicke theory into account. We assume that all of the corrections to General relativity are~small.

After the variable replacing, we rewrite the Equation~(\ref{dotphi}), as follows
\begin{equation}
\frac{d\phi}{dA}=\dot\phi_i\frac{A_i^3}{A^3}\left(\frac{A_{\rm max}^2}{A^2}-1\right)^{1/2}.
\end{equation}

Its solution is
\begin{equation}
\phi=\phi_i+\frac{\dot\phi_i A_i^3}{A_{\rm max}^2}\left(\cot\psi_i-\cot\psi\right).
\label{phisol}
\end{equation}

Substituting expressions for the zero approximation on the right, we get at $\psi=\pi/2$
\begin{equation}
\phi_{\max}=\phi_i+2\dot\phi_it_i.
\label{phimax}
\end{equation}

Here, as~before, the~index ``max'' denotes the values at the moment when the expansion of the disturbed region stops, i.e.,~    $dS/d\tau=0$.

We introduce small corrections to the zero approximation, as follows
\begin{equation}
\phi_i=G^{-1}(1+s_i), \qquad \phi=G^{-1}(1+s_i+p), \qquad S=A(1+\alpha),
\end{equation}
where $s_i$ is the relative deviation of $\phi$ from $G^{-1}$ at the initial moment, $p$ and $\alpha$ are new small functions, of~which the first was already calculated above in Equation (\ref{phisol}). Let us write the Equation~(\ref{meqs}) that is linearised over $\alpha$:
\begin{equation}
2\frac{\dot A}{A}\dot\alpha-2\alpha\frac{1}{A^2}=-\frac{8\pi G}{3}\varepsilon_i S_i^4\frac{1}{A^4}(s_i+p+4\alpha)-\dot\phi_i G S_i^3\frac{\dot A}{A^4}+\frac{\omega}{6}\dot\phi_i^2 G^2S_i^6\frac{1}{A^6}.
\label{lineq}
\end{equation}

The last term contains the small parameter $\dot\phi_it_iG$ squared, but,~because of the possibility of large values of  $\omega$, we keep it. Substituting $\dot A$ from the zero approximation and introducing the notation $x=A^2/A_{\rm max}^2$, we rewrite Equation (\ref{lineq}), as
\begin{equation}
4\frac{d\alpha}{dx}(1-x)-2\alpha=-C_1\frac{s_i+p+4\alpha}{x}-C_2\frac{(1-x)^{1/2}}{x^{3/2}}+C_3\frac{1}{x^2},
\end{equation}
where $C_{1,2,3}$ are some combinations of constants. General solution of the above equation is
\begin{eqnarray}
\alpha&=&\alpha_i\left(\frac{x_i}{x}\right)^{C_1}\left(\frac{1-x}{1-x_i}\right)^{(2C_1-1)/2}+
\nonumber
\\
&+&\frac{1}{4}x^{-C_1}\left(1-x\right)^{(2C_1-1)/2}
\int\limits_{x_i}^{x}dyy^{C_1}\left(1-y\right)^{-(2C_1+1)/2}\left(-C_1\frac{s_i+p}{y}-C_2\frac{(1-y)^{1/2}}{y^{3/2}}+C_3\frac{1}{y^2}\right).
\label{solgen}
\end{eqnarray}

Using the $x $ variable, the~solution Equation (\ref{phisol}) is written, as follows
\begin{equation}
\phi=G^{-1}\left[1+s_i+2\mu-2\mu\delta_i^{1/2}\frac{(1-x)^{1/2}}{x^{1/2}}\right].
\end{equation}

It shows the structure of the function $p$.

From the zero approximation, we have $C_1\approx1$, $C_2\approx2\mu\delta_i^{1/2}$, $C_3\approx2\omega\mu^2\delta_i$, where $\mu=\dot\phi_it_i G$. Using these values, after~integration we obtain
\begin{eqnarray}
\alpha&\simeq&\alpha_i\frac{x_i}{x}\left(\frac{1-x}{1-x_i}\right)^{1/2}
-(\mu+s_i/2-\omega\mu^2\delta_i/3)\left(x^{-1}-x^{-1}(1-x)^{1/2}(1-x_i)^{-1/2}\right)+
\nonumber
\\
&+&\frac{1}{6}\omega\mu^2\delta_ix^{-1}(1-x)^{1/2}\ln\left[\frac{1-(1-x)^{1/2}}{1-(1-x_i)^{1/2}}\frac{1+(1-x_i)^{1/2}}{1+(1-x)^{1/2}}\right].
\label{alppeq}
\end{eqnarray}

At the moment of maximum expansion $x=1$, one gets $\alpha\simeq-\mu-s_i/2+\omega\mu^2\delta_i/3$ from Equation~(\ref{alppeq}). For~verification, let us calculate the same value in a different way. From~the condition $dS/d\tau=0$ in Equation (\ref{meqs}), we obtain an algebraic equation of the fourth order for the value $S_ {\rm max}/S_i$, and~its solution is
\begin{equation}
\left(S_{\rm max}/S_i\right)^2=\frac{4\pi S_i^2\varepsilon_i}{3\phi_{\rm max}}\left(1+\sqrt{1+\Sigma}\right), \quad \Sigma=\frac{3\omega\dot\phi_i^2}{32\pi^2S_i^2\varepsilon_i^2}.
\label{x^2}
\end{equation}

Substituting expressions for the zero approximation, we obtain the same value \mbox{$\alpha\simeq-\mu-s_i/2+\omega\mu^2\delta_i/3$.}

Instead of Equation (\ref{soundeq}), for~the sound wave equation, we have the equation
\begin{equation}
A(1+\alpha)\frac{d\chi}{d\tau}=-\frac{1}{\sqrt{3}}.
\label{newsoundeq}
\end{equation}

Through the variable $x$ it looks, as follows
\begin{equation}
d\chi\simeq-\frac{dx(1-\alpha)}{2\sqrt{3}x^{1/2}(1+x)^{1/2}}.
\label{newsoundeq2}
\end{equation}

Note that, when integrating this equation, the~size of the sound horizon for the first and third terms in Equation (\ref{alppeq}) is accumulated
at early times $\tau$ (small $x$), and,~for the second term, the~size is accumulated at later times, just before the PBH formation. By~performing the integration, we obtain the solution with the correction
\begin{equation}
\chi_{\rm max}\simeq\frac{\pi}{2\sqrt{3}}\left(1+\gamma\right),
\end{equation}
where was denoted
\begin{equation}
\gamma=\frac{1}{\pi}(s_i+2\mu)+\frac{2}{3\pi}\omega\mu^2\delta_i^{1/2}-\frac{2}{\pi}\alpha_i\delta_i^{1/2}.
\label{gameq}
\end{equation}

Instead of Equation (\ref{nonlineq}), we obtain the equation
\begin{equation}
\xi=\sin(\xi^{1/2}(1+\gamma))\sin\left(\frac{\pi}{2}\frac{1}{\sqrt{3}}(1+\gamma)\right).
\label{newnonlineq}
\end{equation}

The corrected solution can be found, as follows
\begin{equation}
\xi=\xi_0+\gamma\left.\frac{d\xi}{d\gamma}\right|_{\gamma=0},
\end{equation}
where the derivative $d\xi/d\gamma|_{\gamma=0}\simeq0.6255$ is calculated by differentiating the Equation~(\ref{newnonlineq}).
By a similar decomposition, we get the final correction for the threshold of PBH formation:
\begin{equation}
\delta_c=\delta_c^{(0)}(1+\lambda\gamma),
\label{deltac1st}
\end{equation}
where $\lambda\simeq0.432$, and~$\gamma$ is given above in Equation (\ref{gameq}). The~expression Equation (\ref{deltac1st}) is the main result of this work. In~the following Sections, we will consider its application for the effect of PBH clustering and, accordingly, for~the rate of gravitational bursts producing by the PBH~merge.

It is nontrivial to choose the initial moment of time $t_i$, since, in Equation (\ref{gameq}), the term in parentheses at the linear stage is invariant with respect to the choice of the initial moment and it is equal to \mbox{$(G\phi_{\rm max}-1)$}, which follows from Equations (\ref{phisol}) and (\ref{phimax}), but~the other terms depend on $t_i$. To~fix the initial conditions, we assume that the moment $t_i$ corresponds to the end of inflation. This model is sufficient for demonstrating the effect of PBH~clustering.

Note that the corrections to the threshold in the Eddington-inspired-Born-Infeld gravity~\cite{Che19} depend on the epoch of PBH formation. However,~our result 
(\ref{gameq}) and (\ref{deltac1st})) is independent of the epoch of collapse. In~Brans-Dicke gravity, the threshold only depends on the initial~conditions.

In this section, we have considered small corrections to General relativity from Brans-Dicke theory. It is interesting to qualitatively discuss what would happen in the case of larger corrections. The~equations for the metric in the static spherically-symmetric case, if~$\dot\phi=0$, was obtained in~\cite{Wei72}  (Equations (9.9.24)--(9.9.26)). In~this case, there is no new characteristic scale, but~only power-law corrections to the post-Newtonian approximation. The~situation is different when $\dot\phi\neq0$.
If,~for~example, the~term $1/S^2$ in Equation (\ref{meqs}) is of the order of the last term, one has the characteristic scale
\begin{equation}
S_D\sim c|\dot\phi/\phi|^{-1}(6/\omega)^{1/2}.
\end{equation}

One could formally relate this scale to the radius at which a flat rotation curve begins to appear in galaxies, which is attributed to dark matter in General relativity. According to astronomical constraints, $\dot G/G<10^{-11}$~yr$^{-1}$. Therefore,
\begin{equation}
S_D>4\times10^2\left(\omega/40000\right)^{-1/2}\mbox{~Mpc},
\end{equation}
where $G\sim\phi^{-1}$. This scale is of the order of large galaxy superclusters scale. In~modified gravity theories~\cite{Mil83-1,Mil83-2,Arr14}, other expressions for the new scale were obtained in terms of parameters of a particular theory and the new scales could be~smaller.


\section{Clustering of Black~Holes}

\subsection{Perturbations in the PBH Number~Density}

To illustrate the effect, let us consider a simple model in which PBHs have a monochromatic mass spectrum, i.e.,~all PBHs are formed with the same mass $M_{\rm PBH}$. We assume also that the perturbations are Gaussian, and~the mean squared value of perturbations on the scale of this mass at the moment of horizon crossing is denoted by $\sigma_{\rm H}$. For~simplicity, let the scalar field have perturbations on a larger scale, which currently contains a mass of cold dark matter $M\gg M_{\rm PBH}$. At~the stage of inflation, perturbations with distinguished scales are generated if the inflationary potential has features, such as local flat segments~\cite{Sta92,IvaNasNov94}.

The PBH is formed under the condition Equation (\ref{usl0}). The~probability of PBH formation, i.e.,~the~fraction of radiation that transformed into the PBHs at the time of their formation, is~written~as
\begin{equation}
\beta(\delta_c)=
\frac{1}{\sqrt{2\pi}\sigma_{\rm H}}
\int\limits_{\delta_c}^{1}d\delta
\exp\left(-\frac{\delta^2}{2\sigma_{\rm H}^2}\right)\simeq
\frac{\sigma_{\rm H}}{\delta_c\sqrt{2\pi}}e^{-\delta_c^2/(2\sigma_{\rm H}^2)},
\label{bet}
\end{equation}
where the last expression is obtained by considering the tail of the Gaussian distribution $\delta\geq\delta_c\gg\sigma_{\rm H}$, and~the upper limit of integration is not important. Nowadays, the cosmological parameter of PBHs
\begin{equation}
\Omega_{\rm PBH}\simeq\beta \frac{a(t_{\rm eq})}{a(t_{\rm max})}.
\label{eqom}
\end{equation}
At the radiation-dominated stage, the~scale factor of the universe is $a(t)\propto t^{1/2}$.

If one assumes that the fraction $\sim$~$1$ of all LIGO/Virgo events is due to PBHs with masses \mbox{$M_{\rm PBH}\sim30M_\odot$}, then the PBHs constitute the fraction $\Omega_{\rm PBH}/\Omega_m\sim10^{-3}$ of all dark matter~\cite{Sasetal16}, where~$\Omega_m$ is the cosmological density parameter of dark matter. From~Equations (\ref{bet}) and (\ref{eqom}), one~obtains $\sigma_{\rm H}\simeq0.013$ in this~case. 

Now, we will show that perturbations in the scalar field that give a correction in Equation (\ref{deltac1st}) are translated into large-scale perturbations in the PBH number density, i.e.,~biasing effect is in place. The inhomogeneities of the scalar field and its time derivative in combination Equation (\ref{gameq}) modulate the~PBH distribution. Moreover, it turns out that perturbations in the PBHs can be significantly amplified in comparison with perturbations in the scalar~field due to the threshold of PBH formation at the tail of the Gaussian distribution.

Substitute the value $\delta_c$ from Equation (\ref{deltac1st}) to Equation (\ref{bet}). Subsequently, the perturbation of the PBH number density is written as
\begin{equation}
\delta_{\rm PBH}[\gamma]=\frac{\beta(\delta_c[\gamma])}{\beta(\delta_c[\gamma=0])}-1\simeq-\frac{(\delta_c^{(0)})^2}{\sigma_{\rm H}^2}\lambda\gamma.
\label{delbh}
\end{equation}

The minus sign is explained by the fact that a positive value of $\gamma$ increases the threshold, and~PBHs form in a smaller number, and,~in the case of a negative $\gamma$, the~opposite situation occurs.
The same is valid for the mean root square values
\begin{equation}
\langle\delta_{\rm PBH}^2\rangle^{1/2}\simeq\frac{(\delta_c^{(0)})^2}{\sigma_{\rm H}^2}\lambda\langle\gamma^2\rangle^{1/2}.
\label{rmsdelbh}
\end{equation}

Because $(\delta_c^{(0)})^2/\sigma_ {\rm H}^2\gg1$, one has $\delta_{\rm PBH}\gg\lambda\gamma$, which is, the~perturbations in the PBH number density are much larger than the initial perturbations in the scalar field. For~the example above ($M_{\rm PBH}\sim30M_\odot$, $\Omega_{\rm PBH}/\Omega_m\sim10^{-3}$), we have $(\delta_c^{(0)})^2/\sigma_ {\rm H}^2\sim10^3$, i.e.,~fluctuations of the scalar field are translated into fluctuations of the PBHs, amplified by three orders of magnitude, according to Equation~(\ref{delbh}). This is the biasing effect in application to the process of PBH formation in the Brans-Dicke~theory.

However, the~PBHs only constitute a small part of all dark matter, so the total value of the density perturbation is
\begin{equation}
\delta_{\rm tot}\simeq-\frac{\Omega_{\rm PBH}}{\Omega_m}\frac{(\delta_c^{(0)})^2}{\sigma_{\rm H}^2}\lambda\gamma\sim-\lambda\gamma,
\label{eqomcorr}
\end{equation}
where $\Omega_{\rm PBH}$ is given by Equation (\ref{eqom}) and~the last estimate corresponds to the example above. Here,~fluctuations in the scalar field are translated into fluctuations in the total density of dark matter, although~with less efficiency than in the PBH component alone, as was in the case Equation (\ref{delbh}).
The clustering effect, which we consider, may also arise in other theories of gravity with additional degrees of freedom, for~example, with~new scalar fields. The~PBH distribution can be modulated by the large-scale inhomogeneities, if~they exist through these degrees of~freedom.

Previously, the~bias effect for PBHs was already considered in General relativity~\cite{Chi06}. The~density perturbation was taken as the sum of the horizon-scale perturbation and long-scale perturbation. Therefore, on~the background of a positive long-scale perturbation, the~total perturbation can more easily overcome the $\delta_c$ threshold than on average. The~usual bias effect applied to galaxies in galaxy clusters also consists in the fact that, in a region with high density, galaxies are formed more likely than in regions with low density~\cite{Kai84,LeeSha98}. In~\cite{Chi06}, a~conformal Newtonian coordinate system was used, in~which density perturbations at the radiation-dominated stage do not evolve and they are equal to perturbations at the moment of the horizon crossing. The~sum of perturbations in conformal Newtonian coordinate system was taken at the single time moment. It was shown later that the effect was overestimated by several orders of magnitude, see~\cite{DesRio18}. The~easiest way to understand the confusion is to use a synchronous reference frame. Indeed, in~the synchronous reference frame $\delta\propto t$ on the super-horizon scales; therefore, at the moment of  PBH formation, the long-wave perturbation has a much smaller values than at the moment of their horizon crossing. Therefore, the~contribution of the long-wave perturbation in the total perturbation is much smaller than it was obtained in the naive calculation in the conformal-Newtonian frame, and therefore,~the biasing effect is very small. A~more accurate method of summation of perturbations at different scales is required for the correct calculation in the conformal-Newtonian frame~\cite{DesRio18}.

\subsection{Influence of Inhomogeneities on the Rate of Gravitational~Bursts}

The number density of PBHs is increased in regions where $\gamma$ has negative sign, see Equation (\ref{delbh}). Let us first consider the case when the clustering of the PBHs is so small that the specified regions did not form virialized clusters up to~now.

Accidentally close PBHs can create a connected pair at the cosmological stage of radiation dominance, which is shrinking in size, due to the emission of gravitational waves and, as~a result, the~PBH merge~\cite{Naketal97}. The~elongation of their orbit and the time of merge is mainly determined by tidal forces from the third nearest PBH. The~effect of inhomogeneities in the distribution of PBHs is not simply increases the probability of pair formation, due to the closer location of PBHs in the inhomogeneities. It also plays a role that the formed pair must merge at the present time in order for the signal to be registered by the detector. This makes the dependence of merge rate on the clustering less obvious. The~rate of PBH merges at the present time $t_0$ was found in~\cite{Naketal97}, as
\begin{equation}
R_b=\left.\frac{\rho_c\Omega_mf}{M_{\rm BH}}\frac{dP(<t)}{dt}\right|_{t=t_0},
\label{rate1}
\end{equation}
where $\rho_c=9.3\times10^{-30}$~g~cm$^{-3}$ is the critical density, $\Omega_m\approx0. 27$, $f$ is the fraction of PBHs in the dark matter, and~the probability that the lifetime of a of PBH pair is less than $t$ has the form
\begin{equation}
P(<t)\sim\left[\frac{37}{29}\left(\frac{t}{t_{\rm max}}\right)^{3/37}-\frac{8}{29}\left(\frac{t}{t_{\rm max}}\right)^{3/8}\right],
\label{pint}
\end{equation}
where
\begin{equation}
t_{\rm max}\sim\frac{5}{512G^3M_{\rm BH}^3}\frac{\bar{x}^4}{f^4},
\label{tmax}
\end{equation}
the average distance between the PBH is
\begin{equation}
\bar{x}=\left(\frac{M_{\rm BH}}{f\rho_{\rm eq}}\right)^{1/3},
\end{equation}
where $\rho_{\rm eq}$ is the average density of dark matter at the moment of transition to the dust-like stage of the universe~evolution.

Because $t_{\rm max}\gg t_0$ is valid for LIGO/Virgo signal parameters, the~first term in Equation (\ref{pint}) gives the dominant contribution to the merge rate. Then for a fixed $f$ we have $R_b\propto \bar{x}^{-12/37}$. Because, for a non-uniform distribution of the PBH $\bar{x}\propto(1+\delta_ {\rm PBH})^{-1/3}$, where $\delta_{\rm PBH}$ is set by Equation (\ref{delbh}), the~final expression $R_b\propto (1+\delta_{\rm PBH})^{-4/37}$ only depends very weakly on $\delta_{\rm PBH}$. Because it seems that the extremely large values of $\lambda\gamma$ cannot be reached, we conclude that the effect of PBH clustering on their collisions in bound pairs formed at the radiation-dominated stage is very small. This may not be in the case when PBHs form virialized clusters, which we will discuss in the next~section.

\subsection{Clusters of Primordial Black~Holes}
\label{pbhclsub}

Perturbations of PBH number density Equation (\ref{delbh}), which arose due to inhomogeneities of the scalar field, and~the total perturbations Equation (\ref{eqomcorr}), belong to the class of entropy perturbations (perturbations with constant curvature). They are independent on the curvature perturbations associated with the perturbations in the relic radiation density. At~the stage of radiation dominance, these entropic perturbations almost do not evolve. Their value increases only 2.5 times due to the Meszaros effect. Additionally,~after the onset of the cosmological stage of matter dominance, these perturbations evolve in the same way as the usual perturbations in dark matter $\propto t^{2/3}$.

Direct collisions of PBHs that are not part of binary systems can occur if the PBHs are located in virialized clusters. Let us consider the formation of such clusters and determine the rate of PBH collisions in~them.

The cross section of the gravitational capture and subsequent collision of two PBHs~\cite{QuiSha87}
\begin{equation}
\sigma_{\rm cap}\approx\frac{3}{2
}\pi
r_{g}^{2}{\left(\frac{c}{v}\right)}^{18/7}, \label{sig18/7}
\end{equation}
where $r_{g}\equiv2GM_{\rm PBH}/c^{2}$, and~$v$ is the relative speed of the two PBHs, and~the light speed was saved in the formulas of this section. The~rate of PBH collisions in a single cluster with a total mass of $M$ (the~sum of the PBH masses and mass of the rest of dark matter)
\vspace{-6pt}
\begin{equation}
\dot N\simeq(1/2)Nn\sigma_{\rm cap}v=9\sqrt2(\Omega_{\rm PBH}/\Omega_m)^2{\left(\frac{v}{c}\right)}^{17/7}\frac{c}{R},
\label{Nrate}
\end{equation}
where, $n$ is the number density of PBHs, $N=M(\Omega_ {\rm PBH}/\Omega_m)/M_{\rm PBH}\simeq(4\pi/3) R^{3}n$ is the number of PBHs in the cluster, $v\sim(GM/R)^{1/2}$, $R$ is the radius of the cluster,
\begin{equation}
R\simeq\left(\frac{3M}{4\pi\rho_{\rm cl}}\right)^{1/3},
\end{equation}
where, according to the spherical model of perturbation evolution, $\rho_{\rm cl}\simeq18\pi^2\bar\rho(t_f)$, $\bar\rho(t)$ is the average cosmological density of dark matter at time $t$, and~$t_f$ is the moment when the perturbation that is evolving according to the law $\propto t^{2/3}$ reaches the value $\simeq2.81$.

The value $\Omega_ {\rm PBH}/\Omega_m\leq10^{-3}$, because~otherwise the rate of gravitational wave bursts due to initially binary systems would be too high (see the previous section). The~rate of gravitational bursts in a unit volume is
\begin{equation}
R_b=\dot N B\frac{\rho_c\Omega_mf_{\rm cl}}{M},
\label{rate2}
\end{equation}
where $B$ is the gain factor due to dynamic effects, as discussed below, and~the fraction of clusters in the dark matter composition
\begin{equation}
f_{\rm cl}=\frac{1}{\sqrt{2\pi}\Delta}
\int\limits_{\delta_{\rm tot}}^{\infty}d\delta
\exp\left(-\frac{\delta^2}{2\Delta^2}\right)=\frac{1}{2}\left[1-{\rm Erf}\left(\frac{\delta_{\rm tot}}{\sqrt{2}\Delta}\right)\right],
\label{fracerf}
\end{equation}
where $\Delta=\langle\delta_{\rm tot}^2\rangle^{1/2}$. The observations give $R_b\sim10$~yr$^{-1}$~Gpc$^{-3}$ \cite{Sasetal16}.

Numerical calculation using the above equations gives the necessary value of perturbation
\begin{equation}
\delta_{\rm tot} f_{\rm cl}^{14/31}\simeq0.1\left(\frac{M}{10^5M_\odot}\right)^{22/93}\left(\frac{R_b}{10\mbox{~yr$^{-1}$~Gpc$^{-3}$}}\right)^{14/31}
\left(\frac{\Omega_{\rm PBH}/\Omega_m}
{0.002}\right)^{-31/28}\left(\frac{B}
{1000}\right)^{-14/31},
\label{itogdel}
\end{equation}
and $f_{\rm cl}<1$. If~we ignore the dynamic effects and put $B\sim1$, we obtain $\delta_ {\rm tot} f_{\rm cl}^{14/31}\sim2$. This means that, even if we take a sufficiently large $\Delta\sim0.1$, it is impossible to obtain clusters of PBH that give the observed rate of gravitational bursts in this~case.

\begin{figure}[H]
\centering
\includegraphics[width=10 cm]{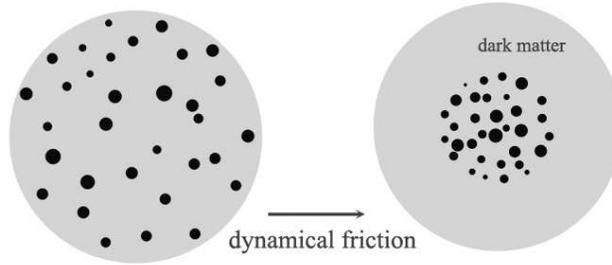}
\caption{A compact primordial black hole (PBH) subsystem is formed inside the object under the influence of dynamical~friction.}
\label{gr2}
\end{figure}

However, there are two dynamical evolutionary effects that can increase the rate of PBH collisions in clusters. The~first effect is the dynamical friction of PBHs in the medium of dark matter in a cluster. Because of dynamical friction, the~PBH lose kinetic energy and settle closer to the center of the cluster, see Figure~\ref{gr2}. As~a result, a~more compact PBH subsystem is formed inside the cluster, which mainly consists of dark matter, and~the rate of merge increases. The~second effect is the dynamical evolution of the PBH cluster due to two-body approaches. This evolution occurs over several dozen relaxation times and also leads to the compression of the central part of PBH cluster and a final increase in the merge~rate.

The objects under consideration have a certain density profiles, which we assume to be the profiles of an isothermal sphere with  density $\propto r^{-2}$. In~this case, the~characteristic PBH velocities are the same at any $r$. But~under the influence of dynamical friction, PBHs lose kinetic energy and their orbit contracts. The~law of compression is estimated as follows
\begin{equation}
r(t)=R\left(1-\frac{t}{t_{\rm df}}\right)^{1/2},
\end{equation}
where the characteristic time
\begin{equation}
t_{\rm df}\sim 0.5t_{\rm eq}\frac{M}{M_{\rm PBH}}\delta_{\rm tot}^{-3/2}.
\end{equation}

It can be seen from Equation (\ref{Nrate}) that the compression of the PBH subsystem leads to a proportional increase in the merge rate. Accordingly, $B$ is equal to the compression ratio $(r(t)/R)^{-1}$. However, the~numerical simulations are required in order to~accurately calculate all of the processes that accompany the cluster compression. The~calculation of this section should be considered as a rough estimate. The~normalizing value $B\sim10^3$ in Equation (\ref{itogdel}) is chosen from the condition that, when the PBH cluster in the isothermal profile $M(r)\propto r$ is compressed by three orders of magnitude in radius, the~PBH gravity in the central region of the object will begin to prevail over the dark matter gravity, since~$\Omega_{\rm PBH}/\Omega_m\sim10^{-3}$. From~the condition $t_{\rm df}<t_0$, we obtain
\begin{equation}
\delta_{\rm tot} >0.044\left(\frac{M}{10^5M_\odot}\right)^{2/3}\left(\frac{M_{\rm PBH}}{30M_\odot}\right)^{-2/3}.
\label{deldf}
\end{equation}

Two-body relaxation leads to dynamical evaporation of the outer part of the PBH cluster and further compression of its central core. In~general, this will cause an additional increase in the rate of PBH merge. Thus, it can be seen that, for $\delta_ {\rm tot}\sim\Delta\sim0.1$, the~conditions are reached under which the rate of PBH merge in clusters begins to prevail over the rate of their merge in the absence of perturbations of scalar field of the Brans-Dicke theory. The~exact calculation of this effect is difficult, due to the presence of complex dynamic evolution of the PBH cluster. The~possibility of reaching $\Delta\sim0.1$, according to Equation (\ref{eqomcorr}), depends on the value of perturbations in the scalar field of the Brans--Dicke~theory.


\section{Discussion}

In this paper, within~the framework of the Brans-Dicke theory, the~PBH formation was studied. A~region of space collapsing into a black hole was modelled as a part of a closed universe on the flat background. Corrections were found to the threshold of PBHs formation, which was previously calculated within the General relativity~\cite{Car75}. It turned out that these corrections depend on the scalar field of the Brans-Dicke theory and on its derivative that was taken at some initial~time.

PBHs are formed in regions with sufficiently large curvature perturbations. If~the scalar field is distributed statistically independent with respect to the distribution of curvature perturbations, then~the perturbations of the scalar field can modulate the number density of PBHs in the universe, which leads to inhomogeneities or clustering of the PBHs. In~the case of strong perturbations of scalar field, we~can even expect the formation of virialized clusters of PBHs. The~collisions of PBHs in clusters occur more frequently than on average, and~this may affect the predicted rate of events for LIGO/Virgo detectors if some of the recorded events are explained by the PBHs. We showed that the effect of PBHs clustering is indeed the case in the Brans-Dicke theory. The~formation of PBHs clusters in a minihalo of dark matter can significantly affect the rate of PBH mergers and the observed rate of LIGO/Virgo gravitational wave~events.

PBH clustering can also influence the formation of structures in the dark matter at sub-galactic scales. This effect may be important in the formation of the first stars and massive black holes in gas clouds in the early universe, since the regions of clustered PBHs create seeds for large-scale density perturbations. An~example of the formation of such an object from the dark matter with a cluster of PBHs in the center was discussed in Section~\ref{pbhclsub}. However, a~detailed study of this effect on a large-scale structure is beyond the scope of this paper. Note that only the results of the Section~\ref{pbhclsub} are applicable to clusters of PBHs that could have been formed due to the other possible mechanisms, and~not only in the Brans-Dicke theory. The~structure and the dynamical evolution of such clusters will be the same. It will be a PBH subsystem immersed in the dark matter halo and compressed due to the dynamic friction. In~addition, in~the outer regions, the~formation of dark matter halos will continue due to the secondary accretion, as~well as the baryonic matter will be captured. In~these objects, the~first stars and supermassive black holes could have~appeared.

Calculations of~\cite{StaYok95} show that, in the Brans-Dicke theory, the~contribution of the entropy (constant curvature) perturbations generated at the inflation stage, to~the total value of perturbations is negligible. In~this paper, we have shown that the contribution can be amplified by about three orders of magnitude. The~presence of such a contribution of entropy perturbations will not contradict the observation of anisotropy of relic radiation, if~the perturbation spectrum in the Brans-Dicke scalar field is only restricted by subgalactic scales. At~the same time, at~larger scales, as observed in cosmic microwave background, the~differences with the standard picture may be below the level that is available for the current~observations.

\begin{acknowledgements}
Authors are grateful to E.O.~Babichev for useful discussions. This research was supported by the CNRS/RFBR Cooperation program for 2018-2020 n. 1985 (France), Russian Foundation for Basic Research (project n. 18-52-15001 CNRS) (Russia) Modified gravity and black holes: consistent models and experimental signatures, and the research program ``Programme national de cosmologie et galaxies'' of the CNRS/INSU. 
\end{acknowledgements}

\end{document}